


\hoffset=-.7truecm
\voffset=-.7truecm
\hsize=17.7truecm
\vsize=23truecm

\baselineskip=12pt plus 1pt minus 1pt
\tolerance=10000
\parskip=0pt
\parindent=15pt

\def\normal{\baselineskip=12pt plus 1pt minus 1pt}

\pageno=1

\newdimen\myhsize   \myhsize=8.55truecm      
\newdimen\myvsize   \myvsize=46truecm      

\newdimen\pagewidth  \newdimen\pageheight
\pagewidth=\hsize  \pageheight=\vsize

\newinsert\margin
\dimen\margin=\maxdimen
\count\margin=0 \skip\margin=0pt

\def\mydate{February 10, 1994}
\def\myname{Yutaka Hosotani}
\def\mytitle{Spontaneous Breakdown of the Lorentz Invariance}

\def\firstheadline{\hbox to \pagewidth{%
    {{\ninerm \hbox{\vtop{ \hsize=2.8truecm \parindent=0pt
           \leftline{Preprint from}
           \leftline{University of Minnesota}     }}}
      \hfil
          {\ninerm  \hbox{\vtop{ \hsize=2.8truecm \parindent=0pt
                \rightline{UMN-TH-1238/94}
                \rightline{\mydate}         }}} }%
       }}
\def\leftheadline{\hbox to \pagewidth{%
      {\it Page \folio\hss \myname\hss\qquad}%
      }}
\def\rightheadline{\hbox to \pagewidth{%
      {\it \qquad \hss\mytitle \hss Page \folio}%
       }}

\def\otherheadline{
   \ifodd\pageno \rightheadline \else \leftheadline \fi}


\def\onepageout#1{\shipout\vbox{
   \offinterlineskip
   \vbox to 1.truecm{    
      \ifnum\pageno=1 \firstheadline \else\otherheadline\fi \vfill}
   \vbox to \pageheight{
     \ifvoid\margin\else
       \rlap{\kern31pc\vbox to0pt{\kern4pt\box\margin \vss}}\fi
    #1
   \boxmaxdepth=\maxdepth} }
 \advancepageno}

\newbox\partialpage
\def\begindoublecolumns{\begingroup
  \output={\global\setbox\partialpage=\vbox{\unvbox255\bigskip}}\eject
  \output={\doublecolumnout} \hsize=\myhsize  \vsize=\myvsize }
\def\enddoublecolumns{\output={\balancecolumns}\eject
  \endgroup \pagegoal=\vsize}

\def\doublecolumnout{\splittopskip=\topskip \splitmaxdepth=\maxdepth
  \dimen1=\pageheight \advance\dimen1 by-\ht\partialpage
  \setbox0=\vsplit255 to\dimen1 \setbox2=\vsplit255 to\dimen1
  \onepageout\pagesofar \unvbox255 \penalty\outputpenalty}
\def\pagesofar{\unvbox\partialpage
  \wd0=\hsize \wd2=\hsize \hbox to\pagewidth{\box0\hfil\box2}}
\def\balancecolumns{\setbox0=\vbox{\unvbox255} \dimen1=\ht0
  \advance\dimen1 by\topskip \advance\dimen1 by-\baselineskip
  \divide\dimen1 by2 \splittopskip=\topskip
  {\vbadness=10000 \loop \global\setbox3=\copy0
    \global\setbox1=\vsplit3 to\dimen1
    \ifdim\ht3>\dimen1 \global\advance\dimen1 by1pt \repeat}
  \setbox0=\vbox to\dimen1{\unvbox1} \setbox2=\vbox to\dimen1{\unvbox3}
  \pagesofar}



\font\tenss=cmss10

\font\bfBig=cmb10  scaled\magstep2

\font\eightrm=cmr8
\font\ninerm=cmr9

\font\nineit=cmti9

\font\ninemit=cmmi9

\font\eightsl=cmsl8

\font\ninebf=cmbx9

\def\myref#1{ [{#1}]}

\def\short#1{\hbox{$\kern .1em {#1} \kern .1em$}}


\def\hspace{~\hskip 1cm ~}

\def\big{\displaystyle \strut }

\def\N{\kappa}
\def\g{{g}}

\def\vp{ {\vec p} }

\def\seq{\hbox{$\,=\,$}}
\def\ZZ{ \hbox{\tenss Z} \kern-.4em \hbox{\tenss Z} }

\def\L{ {\cal L} }

\def\E{ {\cal E} }
\def\O{{\rm O}}

\def\ep{\epsilon}
\def\eps{\varepsilon^{\mu\nu\rho}}
\def\d{\partial}

\def\Tr{ {\,{\rm Tr}\,} }

\def\d{\partial}
\def\la{\raise.16ex\hbox{$\langle$} \, }
\def\ra{\, \raise.16ex\hbox{$\rangle$} }
\def\st{\, \raise.16ex\hbox{$|$} \, }
\def\go{\rightarrow}

\def\det{ {\rm det}\, }
\def\Spect{ {\cal S}\, }
\def\pert{ {\rm p.v.} }
\def\gr{ {\rm g.s.} }

\def\slave{ {l_{\rm ave}^{\,2} } }
\def\ilave{ {l_{\rm ave}^{-1}} }

\def\psibar{ \psi \kern-.65em\raise.6em\hbox{$-$} }
\def\Lbar{ {\cal L} \kern-.65em\raise.6em\hbox{$-$} }


\vglue .5cm

\baselineskip=25pt

\centerline{\bfBig Spontaneous Breakdown of the Lorentz Invariance}

\vskip .3cm

\baselineskip=12pt

\centerline{\ninerm  YUTAKA HOSOTANI}
\centerline{\nineit School of Physics and Astronomy, University of Minnesota,
               Minneapolis, MN 55455}

\vskip .2cm

\centerline{\eightsl Type-set by plain \TeX }

\baselineskip=10pt

\centerline{ \hbox{ \vtop{  \hsize=14truecm
{\eightrm
\midinsert    
We re-examine three-dimensional gauge theory with a
Chern-Simons term in which the Lorentz invariance is spontaneously broken
by dynamical generation of a magnetic field.  A non-vanishing magnetic field
leads, through the Nambu-Goldstone theorem, to the decrease of zero-point
energies of photons, which accounts for a major part of the mechanism.
The asymmetric spectral flow plays an important role.   The instability in pure
Chern-Simons theory is also noted.
 \endinsert
}  }}}


\begindoublecolumns

\vskip .5cm
\normal

In the previous paper\myref{1} we have shown that in a class of
three-dimensional gauge theories described by
$$\eqalign{
\L = &- {1\over 4} \, F_{\mu\nu}F^{\mu\nu} - {\N\over 2} \,
\eps A_\mu \d_\nu A_\rho   \cr
\noalign{\kern 4pt}
&+ \sum_a  {1\over 2} \, \big[ \,\psibar_a \, , \,
  \big( \gamma^\mu_a (i \d_\mu + q_a A_\mu)
    - m_a \big) \psi_a \, \big] ~, \cr}  \eqno(1) $$
the Lorentz invariance is spontaneously broken by dynamical generation of
a magnetic field $B$.  In this paper we shall give additional arguments
in support of this conclusion.

In  (1) $\psi_a$ is a two-component Dirac spinor, and Dirac matrices
are characterized by their signature
$\eta_a \short{=} {i\over 2} \, \Tr \gamma_a^0\gamma_a^1\gamma_a^2 \short{=}
\pm 1$.  In terms of ${\gamma'}^\mu_a \short{=} -\gamma^\mu_a$,  the fermion
part of the Lagrangian takes the same form as the original one except for
the change in the sign of the mass terms  ($m'_a= -m_a$).  Hence
$(\eta_a \short{=}-,m_a \short{>}0)$  is equivalent
to $(\eta_a \short{=}+, -m_a\short{<}0)$.  Both  descriptions are
useful.  The sign is called ``chirality''.
The model is invariant under charge conjugation so that one can take
$q_a \short{>} 0$ without loss of generality.

In the perturbative  vacuum $\la F_{12}(x) \ra \short{=} 0$.
In the previous paper we have shown that in a class of models
a variational ground state,
in which  $\la F_{12}(x) \ra \short{=}   -B \short{\not=} 0$, has a lower
energy
density than the perturbative vacuum, and therefore the Lorentz invariance is
spontaneously broken.

In the variational ground state the energy spectrum of Dirac particles
is characterized by Landau levels:
$$\eqalign{
E &= \left\{ \matrix{
     + \, \ep (\eta B) \cdot \omega_n \qquad (n \ge 0) \cr
     - \, \ep (\eta B) \cdot \omega_n \qquad (n \ge 1) \cr} \right.  \cr}
    \eqno(2)  $$
where $\omega_n =( m_a^2 + 2nq_a|B|)^{1/2}$.
There is asymmetry in the $n\short{=}0$ modes (zero modes).  They exist in
either
positive or negative energy states, or in other words, only for either
particles or anti-particles\myref{2}.  We have considered variational ground
states in which these lowest Landau levels ($n\short{=}0$) are either empty or
completely filled.   Accordingly a filling factor $\nu_a\short{=} 0$ or $1$ is
assigned.
 A variational ground state is denoted as $\Psi_\gr (B, \{ \nu_a \} )$.

Let us fix   Dirac
matrices $\gamma^\mu_a$ for the moment.  To evaluate physical quantities one
may continuosly change the value of the fermion mass $m_a$ from positive to
negative.  Then except for zero-modes ($n\short{=}0$) in the spectrum
(2) all positive (negative) frequencies remain positive (negative).
However, for $\eta B \short{>}0$ for instance, the positive frequency
zero-modes $E\seq\omega_0\seq m_a$ become negative frequency zero-modes.  There
appears crossing in the spectrum.  (See Fig.\ 1.)  Yet physical quantities
must be continuous functions of $m_a$.

Suppose that $\eta_a\short{=} +$ and $\nu_a\seq 0$.  In expanding the Dirac
field
operator $\psi_a(x)$ in terms of Landau level eigenstates with energy
eigenvalues (2),  annihilation  operators $a_{np}$ of particles
(creation operators $b^\dagger_{np}$ of anti-particles) are associated
with positive (negative) frequency eigenstates.  (Here $p$ is an additional
index to {\hglue 3cm}

\vskip 7.5cm

\vfil

\def\smalleta{\hbox{\ninemit\char'21}}

\def\nineneq{ {\ninerm = \kern-1.em\hbox{/}} }

{ \baselineskip=9pt  \ninerm  \noindent
 Fig.\ 1.   The energy spectrum as a function of a mass {\nineit m},
given by  Eq.\ (2) for a positive {\smalleta}{\nineit B} in arbitrary units.
The {\nineit n}=0 mode exhibits crossing in the spectral flow.}

\eject

\noindent
specify eigenstates.)  The fact that $n=0$
positive frequency becomes negative frequency implies that the annihilation
operator $a_{0p}$ is transformed into the creation operator $b^\dagger_{0p}$
under the change of $m_a$.  Hence  $a^\dagger_{0p} a_{0p} \go
b_{0p} b^\dagger_{0p} = 1 - b^\dagger_{0p} b_{0p}$.  In other words
an empty state $\nu_a=0$ with positive $m_a$ becomes a completely filled
state $\nu_a=1$ with negative $m_a$.

Since fermions with $(\eta_a, m_a \short{<}0)$ are equivalent to those with
$(-\eta_a, |m_a| )$, a continuous change $m_a \go - m_a$ results in the
transformation of  $(\eta_a, \nu_a, |m_a|) \go (-\eta_a, 1-\nu_a, |m_a|)$.
Therefore any physical quantities, $R$, must satisfy
$$R(\eta_a, \nu_a, m_a^2) = R( -\eta_a, 1 -\nu_a, m_a^2) ~~.
    \eqno(3)  $$

The charge density $\la j^0 \ra \short{\equiv} J^0$ is a physical quantity.  It
has been shown that $J^0$ is non-vanishing in the presence of a magnetic
field\myref{1,3}:
$$J^0(x) = {1\over 2\pi} \sum_a  \eta_a q_a^2 \,
   (\nu_a - \hbox{${1\over 2}$}) \cdot B  ~, \eqno(4) $$
which satisfies (3).  The factor $\eta_a (\nu_a - {1\over 2})$
reflects the asymmetric spectral flow.

Combined with the Euler equation,  Eq.\ (4) leads to
 a consistency condition for having $B\not= 0$:
$$ \N  = {1\over 2\pi} \sum_a  \eta_a q_a^2 \,
   (\nu_a - \hbox{${1\over 2}$})  ~.  \eqno(5)  $$
With a given  bare Chern-Simons coefficient,
filling factors $\{ \nu_a \}$ are not arbitrary.   It is a necessary condition
for having $B\short{\not=}0$.  It is the purpose of this paper to show that in
a
wide class of models the condition (5) also serves as a
sufficient condition for having $B\short{\not=}0$.

Let us quote some of the  results in ref.\ [1]  relevant for our discussion
below.

To find the difference, $\Delta \E \short{=} \E_\gr \short{-} \E_\pert$,
in the energy densities of the variational ground state and perturbative
vacuum, we split the gauge coupling $q_a A_\mu$ into two parts,
$q_a A^{(0)}_\mu + \alpha q_a A^{(1)}_\mu$, where $A^{(0)}_\mu$ corresponds to
a dynamically generated $B$.  The auxiliary parameter $\alpha$,
ranging from 0 to 1,  has been introduced in the coupling of fluctuations,
defining $\Delta \E(\alpha)$.   Our primary interest is  $\Delta \E(1)$.

It is easy to see that $
\Delta \E(0)\short{=} {1\over 2 } B^2  +  \Delta \E_{\, \rm f}$
where the first term is the Maxwell energy, while the second term represents
the shift in fermion zero-point energies due to a constant magnetic field
$B$.   In the massless fermion limit $\Delta \E_{\, \rm f}$ is positive and is
${\rm O}(|B|^{3/2})$.  At $\alpha\short{=} 0$ there arises no change in
the photon spectrum.

$\Delta\E(1) - \Delta\E(0)$ is found by adiabatically switching $\alpha$ on
from 0 to 1.  It can be expressed in terms of photon propagators:
$$\eqalign{
&\Delta\E(1) - \Delta\E(0) \cr
&=i\int_0^1 {d\alpha\over \alpha} \int {d^3p\over (2\pi)^3} ~
tr \, D_0^{-1}(p) \, \Big\{ D(p)_\gr - D(p)_\pert \Big\}. \cr}
   \eqno(6)  $$
Here $D^{\mu\nu}_0(p)$ is the bare photon propagator, the same in both
variational ground state and perturbative vacuum.  $D^{\mu\nu}(p)_\gr$
and $D^{\mu\nu}(p)_\pert$ are the full photon propagators with a given $\alpha$
in the  variational ground state and pertubative vacuum, respectively.
Notice that Eq.\ (6) is exact.

There are at most three invariant functions $\Pi_k(p^2_0, {\vec p \,}^2)$
($k=1 \sim 3$) to parametrize  the sum of one-particle irreducible diagrams
$\Gamma =D_0^{-1} - D^{-1}$\myref{5,6}.
$$\eqalign{
&\Gamma^{\mu\nu}(p)= ~ (p^\mu p^\nu - p^2 g^{\mu\nu} ) \Pi_0
 + i \ep^{\mu\nu\rho} p_\rho \Pi_1 \cr
&\hskip .2cm +(1-\delta^{\mu 0} ) (1-\delta^{\nu 0})
  (p^\mu p^\nu - {\vec p\,}^2 \delta^{\mu\nu}) (\Pi_2 - \Pi_0) . \cr}
  \eqno(7) $$
Then $\det D^{-1} = (p^2)^2 \cdot \Spect (p)$ where
$$\Spect (p) =  (1+\Pi_0) (p^2 + p_0^2 \Pi_0
 - {\vec p \,}^2\Pi_2 ) - (\N - \Pi_1)^2  ~.   \eqno(8) $$
The  photon spectrum is determined by  $\Spect (p) =0$.

So far all expressions are exact.  At this stage we introduce an
approximation in which all $\Pi_k$'s are evaluated to O($\alpha^2$).  In this
approximation the $\alpha$ integral in (6) can be
readily performed, yielding
$$\Delta\E(1) - \Delta\E(0)
= - {i\over 2} \int {d^3p\over (2\pi)^3} ~
\ln \, {\Spect(p)_\gr \over \Spect(p)_\pert } ~.
   \eqno(9)  $$
The one-loop approximation has been adopted to
$\Pi_k(p)$'s, but not to $\Delta\E$.

$-\Pi_1(p \short{=}0)$ is the induced Chern-Simons coefficient.  It was
evaluated by explicit computations and by employing the
Nambu-Goldstone theorem.  One-loop computations in the variational ground
state have yielded\myref{1}
$$\Pi_1(0)_\gr = {1\over 2\pi} \sum_a  \eta_a q_a^2 \,
   (\nu_a - \hbox{${1\over 2}$})    \eqno(10)  $$
which satisfies (3).
The consistency condtion (5) may be written as
$$\N = \Pi_1(0)_\gr  ~.  \eqno(11)  $$
In the perturbative vacuum\myref{3} one finds
$\Pi_1(0)_\pert \short{=} -\sum_a \eta_a q_a^2 / 4\pi$.

On the other hand the Nambu-Goldstone theorem associated with the spontaneous
breaking of the Lorentz invariance due to a non-vanishing
$-\la F_{12} (0) \ra \short{=} B$ implies that a photon becomes the
Nambu-Goldstone boson in the sense that its spectrum satisfies\myref{7}
$$\lim_{\vp \go 0} p_0(\vp\,) = 0 ~. \eqno(12) $$
Suppose that fermion masses are small, but finite.  Then
 $\Pi_0(0), \Pi_2(0) \short{\not=} 0$.  Since the photon spectrum is given by
$\Spect(p) \short{=}0$ with (8), the Nambu-Goldstone boson nature of a photon
implies the relation (11) to all order.  In other words
the necessary condtion (5) for having $B\short{\not=}0$ is
 also a consequence of $B\short{\not=} 0$.

Now consider a chirally symmetric model consisting of
 $N_{\rm f}$ pairs of $\eta_a\short{=} +$ and $-$ fermions with the same mass
$m_a\short{>} 0$ and  charge $q_a\short{>}0$.   In this model one has $\sum_a
\eta_a q_a^2\short{=} 0$, and
 $\Pi_1(p)_\pert \short{=}0$ exactly in the perturbative vacuum.
We suppose that the bare Chern-Simons
coefficient and  filling factors $\{ \nu_a \}$ of the variational ground state
satisfy the condition (5): $\N \short{=} \sum_a \eta_a \nu_a
q_a^2/2\pi$.

For small $|B|$,  in the  limit $m_a \short{\go} 0$,
$$\eqalign{
\Pi_k(p)_\pert &\sim \Pi_k(p)_\gr   \qquad (k=0,2) \cr
&= \sum_a  {q_a^2 \over  16\sqrt{ -p^2} } + \O (B^2)  \cr
\Pi_1(p)_\pert &= 0 \cr
\Pi_1(p)_\gr &= \sum_a  {\eta_a \nu_a q_a^2\over \pi (2 - p^2 l_a^2)}
   + \O(B^2)    \cr}    \eqno(13) $$
where $ l_a^{-2} \short{=} q_a |B|$.

The shift in the energy density $\Delta\E(1)$ is found by inserting
(13) into (9).   The result is
$$\eqalign{
\Delta\E(1) = &- {\sum \eta_a \nu_a q_a^3 \over 2\pi^3}  \cdot
\tan^{-1} {8 \sum \eta_a \nu_a q_a^2 \over \pi \sum q_a^2}
  \cdot |B|  \cr
& + \O(|B|^{3/2}).  \cr}  \eqno(14)  $$
If the coefficient of the linear term ($\short{\propto}|B|$) is negative,
the energy density is minimized at $B\not= 0$, henceforth the Lorentz
invariance is spontaneously broken.
In the previous paper we  have considered a special case with all
$q_a\short{=}e$,
in which the coefficient is negative.  A wide class of models yield a
negative coefficient.

The decrease of the energy density by a non-vanishing $B$ is understood
in terms of zero-point energies.   Return to (8) and (9).
If $\Pi_0$ and $\Pi_2$ were constant, the $p_0$-integral in (9)
would give the difference between the sums of zero-point energies of photons
in the variational ground state and perturbative vacuum.    In perturbation
theory a
photon is topologically massive with a mass given by the bare Chern-Simons
coefficient $\N$.    In the variational ground state, as explained above,
$B \short{\not=}0$ implies that a photon becomes the Nambu-Goldstone boson
satisfying  (12).  As a momentum $|\vp\,|$ increases, all $\Pi_k$'s
approach to zero.   The crossover takes place, as deduced from (13),
around $|\vp\,|\short{=} \ilave $ where
$${1\over \slave}
   = \bigg| {\sum \eta_a \nu_a q_a^2 / l_a^2
         \over \sum \eta_a \nu_a q_a^2}  \bigg|
 = {|\sum \eta_a \nu_a q_a^3 | \over |\sum \eta_a \nu_a q_a^2|} \cdot |B| ~,
     \eqno(15)  $$
provided that  the denominator  is non-vanishing,  or $\N \short{\not=}0$.

Hence in the variational ground state we have, as a rough estimate,
$p_0 \short{\sim} |{\vp\,}|$ for $|\vp\,| \short{<}  \ilave $, and
$p_0 \short{\sim} ({\vp\,}^2 \short{+} \N^2)^{1/2}$ for
$|\vp\,| \short{>} \ilave$.
 The shift in zero-point energies is thus
$$\eqalign{
\Delta \E &\sim \int^{\ilave}
 {d\vp\over (2\pi)^2} ~
{1\over 2} \bigg\{ \sqrt{ {\vp\,}^2} - \sqrt{ {\vp\,}^2 + \N^2} \bigg\}   \cr
\noalign{\kern 5pt}
&= -{1\over 8\pi} {|\N|\over \slave}
    + \O \Big( {1\over l_{\rm ave}^{\,3}} \Big)  \cr
\noalign{\kern 5pt}
&=  - {|\sum \eta_a \nu_a q_a^3 |\over 16\pi^2} \cdot |B| + \O (|B|^{3/2}) ~.
  \cr}  \eqno(16) $$
Comparing (14) and (16), one finds that in
a typical model the shift in zero-point energies of photons explains
about 50\% of the effect.

What we are observing here is the following self-consistent cycle of
arguments.  Suppose that  a photon is topologically massive in perturbation
theory, i.e. $\N \short{\not=}0$.   (i)  $B \short{\not=}0$ implies the
spontaneous breakdown of the  Lorentz invariance.  (ii)  The photon is the
Nambu-Goldstone boson associated with the spontaneous symmetry breaking.  In
particular,  the energy spectrum of photons is significantly lowered
for a momentum $|\vp\,| \short{<}l^{-1}$ where
$l^{-1} \short{\propto}|B|^{1/2}$.  (iii)   Then zero-point energies of photons
are decreased by an amount $\short{\propto}|B|$ for small $|B|$.  (iv) Hence
the
energy is  minimized at $B \short{\not=}0$.

With this perception at hand we recognize that
 the condition (5) or (11) is not
merely a necessary condition for having $B\short{\not=}0$.  It is  also a
sufficient  condition for lowering the energy density, since it represents
the Nambu-Goldstone boson nature of photons.

A few comments are in order.  First, a photon, as the Nambu-Goldstone boson,
has only one degree of freedom, whereas there are two broken Lorentz-boost
generators, $L_k$ ($k$=1,2).  The mismatch in the numbers is traced back
to the facts that $L_k$  is the first moment of the energy-momentum tensor,
and that a photon is a vector.  A photon couples to both generators.
Secondly, $L_k$ does not commute with the
Hamiltonian.  The Lorentz-boosted variational ground state, $\Psi_\gr'$,  does
not have the same energy density as the original one, $\Psi_\gr$.   Also
in the boosted state the current density $J^k$, as well as $J^0$, is
non-vanishing.  $\Psi_\gr'$ is expected to be a stable state among various
states with a constant non-vanishing current density.

One can generalize the above analysis to the case of pure Chern-Simons
gauge fields.   It is most convenient to consider
$$\L = - {\g \over 4} \, F_{\mu\nu}F^{\mu\nu} - {\N\over 2} \,
\eps A_\mu \d_\nu A_\rho + \cdots    \eqno(17) $$
and take the limit $\g \go 0$ at the end.  The formula (9)
still holds with the replacement
$$\Spect (p) =  (\g+\Pi_0) ( g p^2 + p_0^2 \Pi_0
 - {\vec p \,}^2\Pi_2 ) - (\N - \Pi_1)^2  ~~.   \eqno(18) $$

The $p$-integral in (9) can be easily done as before.
In (14) the coefficient of the linear
term is multiplied by $\g^{-1}$, while the Maxwell energy becomes
${1\over 2} \g B^2$.  Therefore the minimum of the energy density is located
around $|B_{\rm min}|\short{\sim} \g^{-2} \sum\eta_a\nu_aq_a^3$.   This also
follows from the scaling argument.  If one redefines, in (17),
$A' = \g^{1/2} A$ and $q_a' = \g^{-1/2} q_a$, then $|B_{\rm min}'| \short{\sim}
{q'}^3$, or equivalently $|B_{\rm min}|  \short{\sim} \g^{-2} q^3$.

The implication to pure Chern-Simons theory is rather puzzling.  As
$\g \go 0$, $|B_{\rm min}| \go \infty$.  The theory is totally unstable.
At $\g\short{=}0$ with massless fermions the Lagrangian does not contain
any dimensional parameters after redefining $A'\short{=} |\N|^{1/2} A$.
Hence the minimum must occur either at $B'\short{=}0$ or at
$|B'|\short{=}\infty$.  Our analysis indicates the latter.

In the literature it is often said that the effect of gauge interactions in
pure Chern-Simons theory is to merely change the statistics of matter fields,
making them anyons.   As a special case let us consider a model consisting of
only one pair of fermions with the same charge $e$ and mass, but with opposite
chirality.  If $\N\short{=} e^2/2\pi$, the consistency condition (5) is
satisfied.  According to the folklore\myref{8} Dirac particles are supposed to
be transformed to free hard-core bosons with spin either 0 or 1.

Our analysis indicates something more drastic, i.e.\  the gauge interactions
induce the instability.  However,  a reservation must be made that the
$\g \short{\go} 0$ limit corresponds to a strong coupling limit so that the
analysis may need refinement.

In this paper we have shown that  the spontaneous breaking of the Lorentz
invariance in a wide class of models is not only the source, but also
a consequence, of the Nambu-Goldstone theorem.  We shall come back for more
details in separate publications.

\bigskip

\centerline{\bf Acknowledgements}
\bigskip

The author would like to thank Yasuhiro Okada for his enlightening comment
on the importance of zero-point energies of photons in the mechanism.  He
also benefited from discussons with Satoshi Iso, Kiyoshi Higashijima,
Chigak Itoi, Choonkyu Lee, and Taichiro Kugo.  He is grateful to the
National Laboratory for High Energy Physics (KEK) in Japan for its hospitality
where most of this work was done.    This work was supported in part
by the U.S.\ Department of Energy under contract no. DE-AC02-83ER-40105,
and by the Japan Society for the Promotion of Science.

\bigskip

\def\ijmpA#1#2#3{{\nineit Int.\ J.\ Mod.\ Phys.} {\ninebf {A#1}}, #3 (19{#2})}
\def\ijmpB#1#2#3{{\nineit Int.\ J.\ Mod.\ Phys.} {\ninebf {B#1}}, #3 (19{#2})}

\def\mplA#1#2#3{{\nineit Mod.\ Phys.\ Lett.} {\ninebf A{#1}}, #3 (19{#2})}

\def\plB#1#2#3{{\nineit Phys.\ Lett.} {\ninebf {#1}B}, #3 (19{#2})}

\def\np#1#2#3{{\nineit Nucl.\ Phys.} {\ninebf B{#1}}, #3 (19{#2})}
\def\prl#1#2#3{{\nineit Phys.\ Rev.\ Lett.} {\ninebf #1}, #3 (19{#2})}
\def\prA#1#2#3{{\nineit Phys.\ Rev.} {\ninebf A{#1}}, #3 (19{#2})}

\def\prD#1#2#3{{\nineit Phys.\ Rev.} {\ninebf D{#1}}, #3 (19{#2})}

\def\cline{\hfil\noexpand\break  ^^J}

\def\myno#1{\item{[#1]}}

\centerline{\bf References}

\bigskip

\baselineskip=11pt
\ninerm

\myno{1}  Y.\ Hosotani, \plB {319} {93} {332}.

\myno{2}  Y.\ Aharonov and A.\ Casher, \prA {19} {79} {2461}.

\myno{3}  A.N.\ Redlich, \prl {52} {84} {18}; \prD {29} {84} {2366};
          K. Ishikawa, \prl {53} {84} {1615}; \prD {31} {85} {1432}.

\myno{4} See e.g.\ A.A.\ Abrikosov, L.P.\ Gorkov, and I.E.\ Dzyaloshinski,
{\nineit ``Methods of Quantum Field Theory in Statistical Physics''} (Dover
Publications, 1975).

\myno{5}  S.\ Randjbar-Daemi, A.\ Salam, and J.\ Strathdee, \np {340} {90}
   {403}.

\myno{6}  Y.\ Hosotani, \ijmpB {7} {93} {2219}.

\myno{7}  Our picture is different from that of Kovner and Rosenstein who
 attempt to regard a photon  as a Nambu-Goldstone boson associated
with the flux symmetry breaking.  A.\ Kovner and B.\ Rosenstein,
\ijmpA {30} {92} {7419}.

\myno{8}  A.M.\ Polyakov,  \mplA {3} {88} {325};
        S.\ Iso, C.\ Itoi, and H.\ Mukaida,  \np {346} {90} {293};
        \plB {236} {90} {287}.

\enddoublecolumns

\begindoublecolumns

\end